\documentclass[aps, prb, twocolumn, amsmath, amssymb, a4paper, 10pt, superscriptaddress]{revtex4-2}

\usepackage{graphicx}
\usepackage[utf8]{inputenc}
\usepackage{bm}
\usepackage{xcolor}
\usepackage{physics}
\usepackage[colorlinks, citecolor=blue, linkcolor=blue]{hyperref}
\usepackage{siunitx}
\usepackage{xspace}
\usepackage{nicefrac}

\usepackage{sidecap}

\begin{document}

\newcommand{\mJmsq}{\milli\joule\per\square\meter}
\newcommand{\FGT}{Fe\textsubscript{5}GeTe\textsubscript{2}\xspace}
\newcommand{\FtGT}{Fe\textsubscript{3}GeTe\textsubscript{2}\xspace}
\newcommand{\FGaT}{Fe\textsubscript{3}GaTe\textsubscript{2}\xspace}

\title{Room temperature magnetic vortices in the van der Waals magnet \FGT}

\author{Elias Sfeir}
\author{Carolin Schrader}
\author{Florentin Fabre}
\affiliation{Laboratoire Charles Coulomb, Université de Montpellier, CNRS, Montpellier, France}
\author{Jules Courtin}
\author{Céline Vergnaud}
\author{Alain Marty}
\author{Matthieu Jamet}
\author{Frédéric Bonell}
\affiliation{Université Grenoble Alpes, CNRS, CEA, SPINTEC, 38054 Grenoble, France.}
\author{Isabelle Robert-Philip}
\author{Vincent Jacques}
\author{Aurore Finco}
\email{aurore.finco@umontpellier.fr}
\affiliation{Laboratoire Charles Coulomb, Université de Montpellier, CNRS, Montpellier, France}

\begin{abstract}
  We investigate the effect of confinement on the magnetic state of a \SI{12}{\nm}-thick \FGT layer grown by molecular beam epitaxy. We use quantitative scanning NV magnetometry to locally extract the magnetization in rectangular uniformly in-plane magnetized microstructures, showing no enhancement of the Curie temperature compared to magnetization measurements performed before patterning the film, in contrast to previous results obtained on thick \FtGT flakes. Under the application of a weak out-of-plane magnetic field, we observe the stabilization of magnetic vortices at room temperature in micrometric squares. Finally, we highlight the effect of the size of the patterned micro-discs and micro-squares on the stabilization of the vortices using experiments and micromagnetic simulations. Our work thus proposes and demonstrates a way to stabilize non-collinear textures at room temperature in a van der Waals magnets using confinement, although we also show that this approach alone is not successful to enhance the Curie temperature of \FGT significantly above \SI{300}{\K}.
  \end{abstract}

  \date{\today}

  \maketitle

  Among the broad diversity of van der Waals magnets discovered in recent years, very few possess a Curie temperature $T_C$ above room temperature~\cite{wangMagneticGenomeTwoDimensional2022}, which is yet a crucial requirement for technological applications. A part of the research on 2D magnets is therefore devoted to efforts towards increasing their Curie temperature. A first approach is theoretical, with high throughput screening of a large number of possible materials in order to identify the most promising ones~\cite{xinAdvancementsHighThroughputScreening2023}. Experimentally, modifying the composition of materials with a rather high $T_C$ is a promising method. It has been successful for \FtGT: in this material, $T_C \simeq \SI{230}{\K}$ in the bulk~\cite{deiserothFe3GeTe2Ni3GeTe2Two2006}, but replacing Ge with Ga leads to an increase of $T_C$ up to 350-\SI{380}{\K}~\cite{zhangRoomtemperatureChiralSkyrmion2024} and to the observation of skyrmions in thick flakes of \FGaT~\cite{liuMagneticSkyrmionsRoom2024, luoManipulationMagneticSpin2025a}. Recent theoretical investigations attribute this effect to higher-order exchange coefficients which are positive in \FGaT in contrast to \FtGT in which they are negative~\cite{kimWhyFe3GaTe2Has2025}. Electrostatic doping~\cite{dengGatetunableRoomtemperatureFerromagnetism2018} or the growth of \FtGT on a topological insulator~\cite{wangRoomTemperatureFerromagnetismWaferScale2020} are also possible solutions to increase its Curie temperature.

   Besides doping or modifying the material itself, another possible approach is to pattern the \FtGT layer in order to benefit from confinement effects. X-ray magnetic circular dichroism and photoelectron emission microscopy (XMCD-PEEM) experiments on \FtGT microstructures have shown a significant enhancement of $T_C$ as well as the stabilization of magnetic vortices~\cite{liPatterningInducedFerromagnetismFe3GeTe22018}. However, in this previous work, the patterning of the microstructures had been achieved using a Ga beam. Ga ions have probably been integrated in the film during the etching process, which could have impacted $T_C$~\cite{yuanModulatingAboveroomtemperatureMagnetism2023}. Further investigations would thus be required to properly determine the influence of confinement on the Curie temperature of \FtGT.

   \begin{figure}[!h]
    \centering
    \includegraphics[scale=1.02]{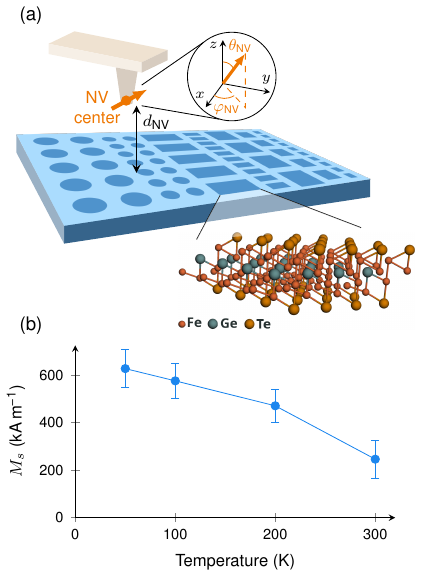}
    \caption{(a) Sketch of the experiment. We use a diamond scanning probe to perform NV magnetometry experiments on a microstructured \FGT film. The top inset defines the angles $\theta_\text{NV}$ and $\varphi_\text{NV}$ which describe the orientation of the NV center. The bottom inset displays the structure of a monolayer of \FGT, drawn using the pyXtal library~\cite{fredericksPyXtalPythonLibrary2021} and a cif file from the Computational 2D Materials Database~\cite{haastrupComputational2DMaterials2018, gjerdingRecentProgressComputational2021}. (b) SQUID measurement of $M_s$, indicating a Curie temperature above \SI{300}{\K} for the full film before patterning.}
    \label{fig:exp}
  \end{figure}

   \begin{figure*}
    \centering
    \includegraphics[scale=0.95]{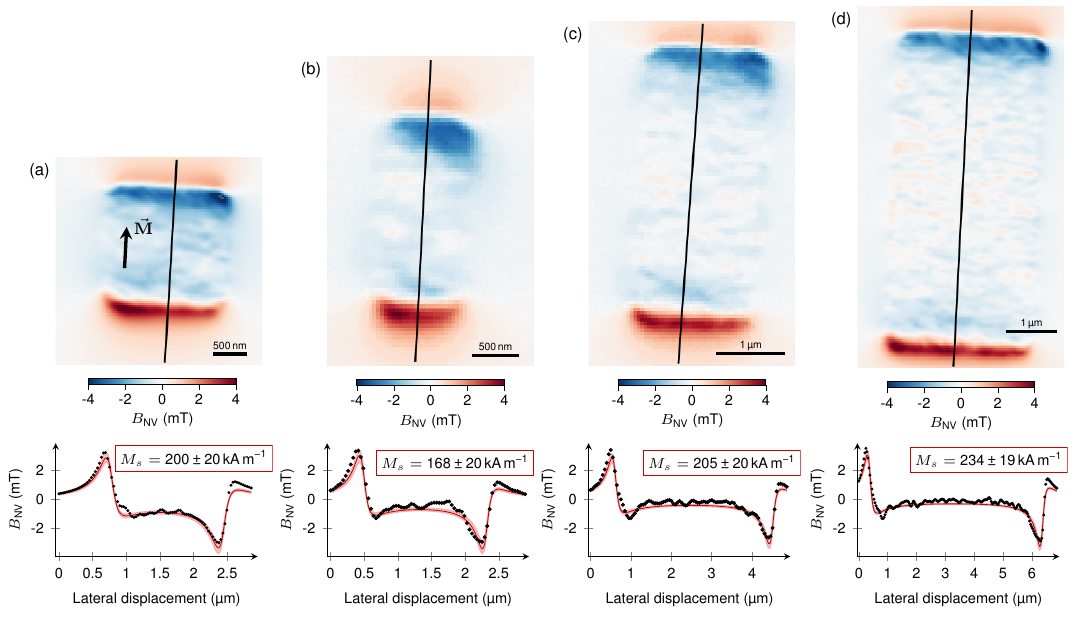}
    \caption{Measurement of $M_s$ in microstructures of different sizes. The maps show the measured stray magnetic field under a tilted external magnetic field of \SI{5.4}{\milli\tesla}, allowing us to stabilize a single domain in each rectangle. The plots below each map show the profiles extracted along the black lines and the corresponding fits with Eqs.~\ref{eq:B_NV} and~\ref{eq:edge_field}, together with the extracted $M_s$ value.}
    \label{fig:Ms}
  \end{figure*}
  
  Increasing the amount of Fe in the material helps as well, as \FGT exhibits a $T_C$ of \SI{310}{\K} in the bulk~\cite{mayFerromagnetismRoomTemperature2019}, and slightly lower but about room temperature in flakes~\cite{mayFerromagnetismRoomTemperature2019} and ultrathin films~\cite{ribeiroLargescaleEpitaxyTwodimensional2022}. Doping \FGT with Co allows reaching an even higher Curie temperature, about \SI{325}{\K}~\cite{zhangEnhancedMagneticElectrical2024}. \FGT is a particularly interesting material, as several groups have reported the finding or the prediction of magnetic bubbles, merons and skyrmions~\cite{gaoManipulationTopologicalSpin2022, schmittSkyrmionicSpinStructures2022a, casasCoexistenceMeronsSkyrmions2023a, gopiThicknessTunableZoologyMagnetic2024, lvManipulatingMagneticBubbles2024, liPredictionStableNanoscale2024}, with discussions about the presence or not of a sizable Dzyaloshinkii-Moriya interaction.

  Here, we investigate the influence of patterning on the Curie temperature of a \FGT thin film grown by Molecular Beam Epitaxy (MBE)~\cite{ribeiroLargescaleEpitaxyTwodimensional2022}, which is thus already ferromagnetic at room temperature. We also examine the effect of confinement on the stabilization of non-collinear magnetic textures, in particular vortices. In order to perform this analysis and be able to extract locally the value of the magnetization in different microstructures, we use scanning NV center magnetometry~\cite{rondinMagnetometryNitrogenvacancyDefects2014}. This technique allows us to map quantitatively the magnetic stray field produced by the \FGT microstructures, non perturbatively and under ambient conditions. All the measurements presented in this work have been performed with a commercial scanning NV center microscope (ProteusQ, Qnami).

  Fig.~\ref{fig:exp}(a) presents a sketch of the experiment. Our sample is a  \SI{11.8}{\nm}-thick \FGT film grown by MBE on Al\textsubscript{2}O\textsubscript{3} and protected with a \SI{3}{\nm}-thick Al capping layer. Details of the growth and of the macroscopic characterization of the film are presented in ref.~\cite{ribeiroLargescaleEpitaxyTwodimensional2022}. The film has been subsequently patterned with electron beam lithography and Ar ion etching in order to create microstructures of various sizes and shapes (squares, rectangles and discs). Macroscopic characterization of the film before patterning using Superconducting Quantum Interference Device (SQUID) magnetometry is presented in Fig.~\ref{fig:exp}(b) and indicates that the film is magnetized in-plane and exhibits a Curie temperature slightly above room temperature, with a saturation magnetization $M_s = \SI{246 \pm 80}{\kilo\ampere\per\meter}$ at \SI{300}{\K}.

  In a first series of measurements presented in Fig.~\ref{fig:Ms}, we perform scanning NV center magnetometry on microstructures of different sizes. We use a permanent magnet to apply a tilted external magnetic field of \SI{5.4}{\milli\tesla} roughly aligned with the quantization axis of the NV center, which is described by the angles $\theta_\text{NV} \simeq \ang{125}$ and $\varphi_\text{NV} \simeq \ang{87}$ defined in Fig.~\ref{fig:exp}(a). The field is sufficient to obtain rectangular microstructures hosting a single uniform magnetic domain, and its out-of-plane component is low enough to avoid tilting the magnetization $\vec{M}$ out of the $ab$ plane. In this case, we expect to detect magnetic stray field at the edges orthogonal to $\vec{M}$, and no field at the edges parallel to $\vec{M}$. This corresponds perfectly to the experimental data shown in Fig.~\ref{fig:Ms} for 4 different rectangular microstructures. 

  Since our measurements are quantitative, we can compute the saturation magnetization $M_s$ from each map, by fitting the stray field to its analytical expression. To achieve this, we first extract a stray field profile across the edges which produce stray field. The position of these line profiles are indicated by the black lines on the images in Fig.~\ref{fig:Ms}. With scanning NV center magnetometry, we measure the component $B_\text{NV}$ of the stray field, which is its projection along the quantization axis of the NV center and can therefore be written as:
  \begin{align}
    \label{eq:B_NV}
    B_\text{NV} = & \ \sin \theta_\text{NV} \cos \varphi_\text{NV} \ B_x + \sin \theta_\text{NV} \sin \varphi_\text{NV} \ B_y \nonumber \\ & + \cos \theta_\text{NV} \ B_z  
  \end{align}
  The magnetic stray field produced at an edge parallel to the $x$ axis and located at $y=y_0$, with $\vec{M}$ orthogonal to the edge and measured at a height $d_\text{NV}$ can be written as~\cite{suppl}:
  \begin{equation}
    \label{eq:edge_field}
      \begin{aligned}
        B_x(y)  = & \  0  \\
        B_y(y)  =  & \ \frac{\mu_0 M_s}{2 \pi} \left[ \arctan\left( \frac{d_\text{NV}-\frac{t}{2}}{y-y_0} \right) - \arctan\left( \frac{d_\text{NV}+\frac{t}{2}}{y-y_0} \right)\right]  \\
        B_z(y)  = & \ \frac{\mu_0 M_s}{4 \pi} \left[ \log\left( (d_\text{NV}-\frac{t}{2})^2 + (y-y_0)^2 \right) \right. \\ & \hspace*{0.9cm} \left. - \log\left( (d_\text{NV}+\frac{t}{2})^2 + (y-y_0)^2  \right)\right]
  \end{aligned}
  \end{equation}
where $t$ is the film thickness. We can fit the experimental line profiles to a combination of Eqs.~\ref{eq:B_NV} and~\ref{eq:edge_field}, considering two opposite edges in each profile. Although the distance $d_\text{NV}$ between the NV center and the surface as well as the angles $\theta_\text{NV}$ and $\varphi_\text{NV}$ can be calibrated independently, we consider $d_\text{NV}$ as a free parameter in the fits because of the presence of resist residues on the surface of the sample. These fits are plotted in the graphs at the bottom of Fig~\ref{fig:Ms}, where the resulting value of $M_s$ is also indicated. The main conclusion from this analysis is that $M_s$ does not vary significantly with the size of the microstructure, and since the average value that we obtain, $M_s = \SI{202 \pm 12}{\kilo\ampere\per\m}$, is in good agreement than the value of $M_s$ measured by SQUID at \SI{300}{\K}, we conclude that the microstructuring does not help increasing $T_C$ in our experiment.

\begin{figure}
    \centering
    \includegraphics[scale=0.95]{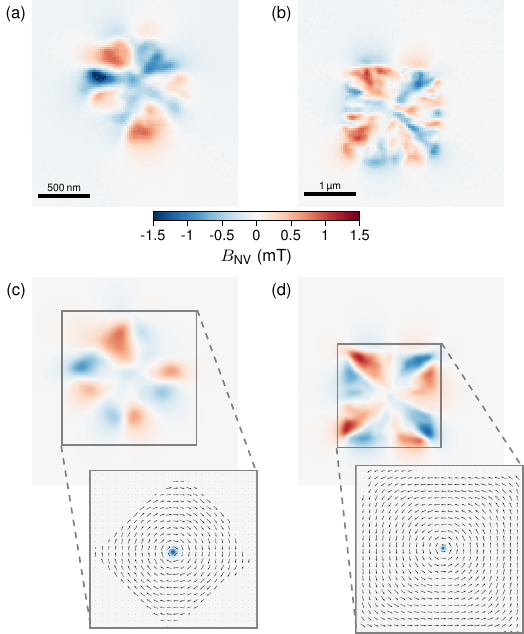}
    \caption{(a)-(b) Measured stray field maps of vortices in square microstructures.  (c)-(d) Computed stray field maps resulting from the magnetic state shown in the insets. The insets show the magnetization configurations obtained from micromagnetic calculations with the parameters from Table~\ref{tab:sim_params}, with the blue color indicating the core of the vortices where the magnetization tilts out-of-plane.}
    \label{fig:vortices}
  \end{figure}

In a second series of measurements, we reduce the applied magnetic field to \SI{3.6}{\milli\tesla} out-of-plane. This field allows us to determine the sign of the magnetic field in the scanning NV magnetometry scans~\cite{fincoSingleSpinMagnetometry2023} and has a limited effect on the magnetic texture inside the \FGT microstructures. Similarly to what was observed in the MFM images of the \FtGT microstructures in ref.~\cite{liPatterningInducedFerromagnetismFe3GeTe22018}, our magnetic stray field maps reveal the presence of magnetic vortices in some of the square structures, see Figs~\ref{fig:vortices}(a)-(b). Indeed, we detect a flower-shaped stray field pattern, which is typical for vortices~\cite{rondinStrayfieldImagingMagnetic2013}, in squares of different sizes and orientations with respect to the crystal structure of \FGT. 

However, as the vortex cores are not visible in our images, we performed additional micromagnetic simulations to really ensure that we are observing vortices. We used the Python package Ubermag~\cite{begUbermagMoreEffective2021} with OOMMF~\cite{oommf} to compute the relaxed magnetic configurations in the two squares considered in Fig.~\ref{fig:vortices}, and computed the associated magnetic stray field map at the NV center height using a scalar potential method~\cite{abertFastFiniteDifferenceMethod2012}. We extracted the shape of the microstructures from the topography measurements, and used the parameters gathered in Table~\ref{tab:sim_params}. We chose a value for the exchange stiffness which corresponds to the order of magnitude given in the literature, \SI{10}{\pico\J\per\meter}~\cite{schmittSkyrmionicSpinStructures2022a, lvManipulatingMagneticBubbles2024, gopiThicknessTunableZoologyMagnetic2024}. Macroscopic measurements of the uniaxial anisotropy $K_u$ shown in Fig.~S3~(a)~\cite{suppl} indicate that it is negative and vanishes around room temperature. Using the slightly positive value of $\SI{18 \pm 30}{\kilo\joule\per\cubic\meter}$ measured at \SI{300}{\K} results in the presence of a large and clearly visible vortex core in the center of the square microstructure (see Fig.~S3), which does not match with our experiments. We have therefore varied $K_u$ according to the error bar in order to get images which are in better agreement with experimental data, and used $K_u = \SI{-20}{\kilo\joule\per\cubic\meter}$. Using this value, the vortex core cannot be distinguished in the stray field images, in agreement with our measurements. We also added some disorder in the simulations in order to mimick local variations of thickness or of $M_s$ by defining random grains with a Voronoi tesselation and attributing them a random $M_s$ value, with a variation of $\pm 20\%$ around the measured mean value $M_s = \SI{202}{\kilo\ampere\per\meter}$. Note that this $M_s$ distribution is not directly related to the measurement uncertainty on the $M_s$ value in Fig.~\ref{fig:Ms}, which provides an average value on the microstructure, but describes very local variations resulting from disorder and thickness fluctuations. 
The NV-to-sample distance $d_\text{NV}$ was assumed rather large, \SI{120}{\nm}, in agreement with the fits shown in Fig.~\ref{fig:Ms} (see actual values in Supplementary Material~\cite{suppl}), and as expected higher than measured in the calibration. The NV orientation $\theta_\text{NV}, \varphi_\text{NV}$ was determined from the calibration measurement (Fig.~S2~\cite{suppl}).
Using this $K_u$ value and including disorder, we obtain stray field maps in good agreement with the experimental ones. 
We therefore conclude that we are actually observing vortices in square \FGT microstructures at room temperature.

  \begin{table}
    \centering
    \begin{tabular}{ll}
      \hline
      Parameter & Value \\
      \hline
      NV height $d_\text{NV}$ & \SI{120}{\nm} \\
      NV polar angle $\theta_\text{NV}$ & \ang{125} \\
      NV azimuthal angle  $\varphi_\text{NV}$ & \ang{87} \\
      Exchange stiffness $A$ & \SI{10}{\pico\joule\per\meter}\\
      Uniaxial anisotropy $K_u$ & \SI{-0.02}{\joule\per\cubic\centi\meter}\\
      Film thickness $t$ & \SI{11.8}{\nm}\\
      Saturation magnetization $M_s$ & \SI{202}{\kilo\ampere\per\meter}\\
      $M_s$ disorder, patch size  &  $\sim \SI{50}{\nm}$\\
      $M_s$ disorder, variation amplitude \hspace*{5mm} &  $\pm$ 20\%\\
      \hline
    \end{tabular}
    \caption{Parameters used in the micromagnetic simulations and the subsequent stray field calculations.}
    \label{tab:sim_params}
  \end{table}

  We performed similar experiments in disc-shaped microstructures, and observed either magnetic single domains in very small discs (diameter below \SI{500}{\nm}), or complicated flower-shaped patterns most probably resulting from disorder (see Fig.~S4~\cite{suppl}). We cannot really conclude experimentally about the presence of vortices in the discs, as only the vortex core is expected to generate stray field in a perfect disc~\cite{shinjoMagneticVortexCore2000a, TetienneQuantitativeStrayField2013}. Disorder has a strong impact on the detected stray field, and our simulations including disorder and a vortex in the center of the discs show a reasonable agreement with the experimental data. 

\begin{SCfigure*}
    \centering
    \includegraphics[scale=1.0]{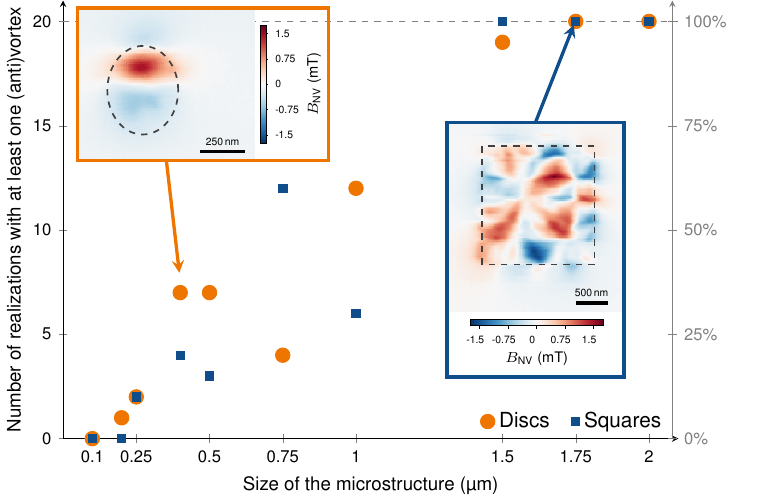}
    \caption{Graph gathering the results of micromagnetic simulations of the magnetic state in discs and squares of various sizes, in the absence of external magnetic field and starting from random configurations. For each size and shape, 20 disorder configurations were considered. In the larger structures ($> \SI{1.5}{\um}$), several vortices and/or anti-vortices were systematically found. The insets show experimental data of a monodomain disc of diameter \SI{0.4}{\um} and a multi-vortex square of size \SI{1.75}{\um}, measured under small out-of-plane magnetic field in order to bias the NV center.\vspace*{10mm}}
    \label{fig:size_effect}
  \end{SCfigure*}

  Our investigation of microstructures of different sizes (\SI{500}{\nm}, \SI{1}{\um}, \SI{2}{\um} and even rectangles of larger dimensions) indicates that there is a higher probability to observe one or several vortices in the large structures than in the small ones. In particular, we found several discs of diameter below \SI{500}{\nm} without any vortex, like the one shown in the left inset of Fig.~\ref{fig:size_effect}, while the large rectangles like the one shown in Fig~\ref{fig:Ms}(d) usually exhibit several vortices in low out-of-plane field. We investigated this question further using micromagnetic simulations, without disorder and external magnetic field applied. We considered squares of various sizes, from \SI{100}{\nm} to \SI{2}{\um} as well as discs with diameters ranging from \SI{100}{\nm} to \SI{2}{\um}. As an initial state, we use a completely random orientation of the magnetization in each cell. We performed the simulation with 20 different random initial configurations for each size, and counted the number of realizations in which one or several vortices or antivortices are present at the end of the relaxation. Our results are gathered in Fig.~\ref{fig:size_effect}.   The same trend is observed for the discs and the squares. In the very small structures of typical size \SI{100}{\nm}, we never find a vortex. When the size increases, the probability to observe a vortex increases, and for a size above \SI{1.5}{\um}, we systematically end up with a complicated magnetic state consisting in several vortices and/or antivortices.

  This size effect has been studied previously experimentally, analytically and numerically~\cite{cowburnSingleDomainCircularNanomagnets1999, schneiderStabilityMagneticVortices2002, metlovStabilityMagneticVortex2002, mejia-lopezDevelopmentVortexState2010, rochaDiagramVortexFormation2010, barmanMagnetizationDynamicsNanoscale2020} in discs and squares of soft ferromagnets. The critical disc radius separating the vortex state from a roughly uniformly in-plane magnetized state in microdiscs depends on the exchange stiffness, the saturation magnetization and on the aspect ratio of the discs. In particular, most of the analytical calculations~\cite{metlovStabilityMagneticVortex2002} have been performed for discs with a thickness larger than the exchange length~\cite{aboDefinitionMagneticExchange2013}:
  \begin{equation}
    \label{eq:lex}
    l_\text{ex} = \sqrt{\frac{2A}{\mu_0 M_s^2}}
  \end{equation}  
  In our case, $l_\text{ex} \simeq \SI{20}{\nano\meter}$, and therefore our discs are significantly thinner than $l_\text{ex}$, meaning that a 2D model is more relevant~\cite{rochaDiagramVortexFormation2010}. It appears that for very thin films, the critical radius allowing the stabilization of vortices in a disc becomes very large~\cite{barmanMagnetizationDynamicsNanoscale2020}, in agreement with our results presented in Fig.~\ref{fig:size_effect}. We also note that the diameter of the core of the vortices in our simulations, which we obtained by fitting a two-dimensional gaussian function to the simulated maps of $M_z$, lies between 60 and \SI{55}{\nm}, again much larger than $l_\text{ex}$. Comparing this large core size with the dimensions of our simulated discs and squares, it appears that, as shown in Fig.~\ref{fig:size_effect}, the \SI{100}{\nm} structures are too small to allow for the proper stabilization of a vortex.

To conclude, using quantitative scanning NV center magnetometry, we demonstrated that microstructuring does not significantly affect the Curie temperature in MBE-grown \FGT ultrathin films. However, by comparing experimental data with micromagnetic simulations, we revealed the presence of magnetic vortices in microstructures. We thus demonstrated here a way to control the formation of non-collinear magnetic textures at room temperature in \FGT, opening perspectives to study the rich physics and potential applications of these objects in the context of 2D and van der Waals materials.   

  \vspace*{11mm}
  
The data that support this work, including the Jupyter notebooks used for the micromagnetic simulations, are available in Zenodo with the DOI 10.5281/zenodo.15805196.\\
  
  \begin{acknowledgments}
   
    The French National Research Agency (ANR) is acknowledged for its support under the program ESR/EQUIPEX+, Grant No. ANR-21-ESRE-0025 (2D-MAG) and the project ELMAX (ANR-20-CE24-0015). For the purpose of Open Access, a CC-BY public copyright licence has been applied by the authors to the present document and will be applied to all subsequent versions up to the Author Accepted Manuscript arising from this submission. 
  \end{acknowledgments}

 
%

  \end{document}